\documentclass[pre,floatfix,twocolumn]{revtex4-1} 
\usepackage[pdfpagemode=UseNone,pdfstartview=FitH,colorlinks=true]{hyperref}
\usepackage{enumitem}
\usepackage{url}
\setlistdepth{9}
\usepackage{graphicx}
\usepackage{epstopdf}
\usepackage{amsmath}
\usepackage{amssymb}
\usepackage{comment}

\renewcommand{\[}{\begin{equation}}
\renewcommand{\]}{\end{equation}}
\newcommand{\bx}{{\bf{x}}}

\newcommand{\bF}{{\bf{F}}}

\begin{document}
\title{State-to-State Cosmology: a new view on the cosmological arrow of time and the past hypothesis}

\author{J. M. Deutsch}
\email{josh@ucsc.edu}
\affiliation{Department of Physics, University of California, Santa Cruz CA 95064}

\author{Anthony Aguirre}
\email{anaguirr@ucsc.edu}
\affiliation{Department of Physics, University of California, Santa Cruz CA 95064}

\begin{abstract}

Cosmological boundary conditions for particles and fields are often discussed as a Cauchy problem, in which configurations and conjugate momenta are specified on an ``initial'' time slice.  But this is not the only way to specify boundary conditions, and indeed in action-principle formulations we often specify configurations at two times and consider trajectories joining them.
Here, we consider a classical system of particles interacting with short range two body interactions, with boundary conditions
on the particles' positions for an initial and a final time. For a large number of particles that are randomly arranged into a 
dilute gas, we find that a {\em typical} system trajectory will 
spontaneously collapse into a small region of space, close to the maximum density that is
obtainable, before expanding out again. If generalizeable, this has important implications for the cosmological arrow of time, potentially allowing a scenario in which both boundary conditions are generic and {\em also} a low-entropy state ``initial'' state of the universe naturally occurs.
\end{abstract}
\maketitle
 
\section{Introduction}

One of the most significant unsolved problems in physics and cosmology is a comprehensive understanding of the ``arrows of time'': many physical processes such as entropy increase, dissipation, records, and causal relations, have a directionality in time even while the laws of gravity and quantum field theory governing particles and fields admit a symmetry including time-reversal.  It is widely believed that most of these arrows are tied to arguments in statistical mechanics: like entropy increase, they relate to a coarse-graining of the underlying microphysical dynamics such that the drive toward the ``future'' corresponds to evolution into more {\em generic} (higher entropy) coarse-grained states. The arrows of time can then be accounted for by postulating that at some time in the (by definition) past, the world was in a state of low entropy; this is often known as the ``Past Hypothesis''~\cite{albert2001time}.

The Past Hypothesis, however, presents a profound conundrum cosmologically. If the arrows of time are to hold throughout cosmic history, the low-entropy state must pertain to the earliest stage of the universe, i.e. to initial cosmological boundary conditions. But it then states that, in effect, the cosmological boundary conditions are enormously {\em special}. Put another way, entropy can be considered the lack of information, and vice-versa.  So the past hypothesis is inserting a vast quantity of information into the cosmological model.  The big-bang cosmology for example is very economical to specify, with just a couple of dozen properties and numbers (e.g.~\cite{tegmark2006dimensionless}); but these would then be supplemented by a vast quantity (potentially of order $10^{122}$ bits or more) of additional implicitly specified information.

The question of the cosmological arrow of time has led to consideration of myriad different ways of specifying cosmological boundary conditions that differ from the ``canonical'' notion of imposing Cauchy initial conditions at some very early time in a big-bang spacetime, just prior to which is a cosmological singularity.  The Steady-State cosmology of the 1960s provided an alternative in which specification of the universe could be done (at a statistical level) at any time, all times being equivalent. Also in the 1960s Gold proposed that a closed, recollapsing universe could (or perhaps would) have low entropy at early {\em and} late times, an idea which was later taken up and then repudiated by Hawking~\footnote{See~\cite{price1997time} for history and references.}. A surprisingly long time later, Aguirre \& Gratton~\cite{aguirre2002steady,aguirre2003inflation}, Carroll \& Chen~\cite{carroll2004spontaneous}, and later Barbour et al.~\cite{barbour2014identification} investigated the idea that a non-singular low entropy boundary condition would lead to a time-symmetric universe with the arrow of time ``switching'' at the boundary surface.  A variation on this idea goes back to Cocke~\cite{cocke1967statistical}, which which one postulates two high entropy states bounding a history of decreasing and then increasing entropy. The  framework used was stochastic transitions to states of lower entropy that
are increasingly less probable. This makes the probability of such evolution an exponentially small subset of possible trajectories.\footnote{
This sort of global fluctuation was shown~\cite{aguirre2012out} to possess time symmetry
where the collapsing and expanding phase would be statistically identical to each other.} 

Here we examine two time boundary conditions, that is
{\em high}-entropy (generic) boundary conditions at {\em two} times,
but from a different point of view. As had been suggested earlier~\cite{cocke1967statistical},
one would imagine this being quite boring, because intuitively the
entropy should stay high in-between, leading to an equilibrium
cosmology with just a tiny chance of a low-entropy state occurring in between.  We argue that, remarkably, this is {\em not} the case,
at least in classical mechanics: specifying a generic ``early'' and
``late'' state leads predominantly to trajectories between those states that looks
low-entropy in the middle.  
Extrapolated to larger scales and
additional physics, this could lead to a new and quite different
perspective on cosmological boundary conditions and the arrows of
time.

\begin{comment}
A variation that has not to our knowledge been explored is to pose {\em high}-entropy (generic) boundary conditions at {\em two} times.  One would imagine this being quite boring, because intuitively the entropy should stay high in-between, leaving to an equilibrium cosmology.  We argue that, remarkably, this is {\em not} the case, at least in classical mechanics: specifying a generic ``early'' and ``late'' state leads to a trajectory between those states that looks low-entropy in the middle.  Extrapolated to larger scales and additional physics, this could lead to a new and quite different perspective on cosmological boundary conditions and the arrows of time.
\end{comment}

In Sec.~\ref{sec-BCs} we discuss how to place classical mechanical boundary conditions at two different times (a highly unexplored issue, as it turns out).  Then in Sec.~\ref{sec-model} we specify the physical model. Results of numerical work (with methodological details given in Appendix \ref{app:num_method}) are shown in Sec.~\ref{sec-num}, then Sec.~\ref{sec-largeN} gives analytic arguments as to how we expect these results to generalize to more particles.  We then discuss both implications and extensions to other physics in Sec.~\ref{sec-discuss}.

\section{Two-time boundary conditions}
\label{sec-BCs}

The vast majority of work in physics starts with the presumption that a system evolves as specified by boundary conditions placed at a particular time. In classical calculations, information given about initial positions and momenta 
is then used to calculate subsequent evolution. But this does not have to be the only paradigm for
understanding the physical world. Here we ask what happens if instead we are given
particle positions at both an initial and a final time, but are not given information about momenta.

The mathematical formulation of classical mechanics in terms of initial and final states is indeed quite standard:
this is precisely what is used in the definition of the action.  The Stationary Action Principle
allows one to obtain all paths that solve Newton's laws under the condition of fixed end points.
This also connects nicely to the path integral formulation of quantum mechanics, where the propagator is
determined by doing a weighted sum over all paths, but with fixed end points, and in this framework the classical trajectories can be identified as those naturally emerging from the $\hbar \rightarrow 0$ limit.

Despite its mathematical elegance, there has been little work in understanding the characteristics
of such classical paths within the framework of statistical mechanics, where one is most often interested
in a large number of particles and chaotic dynamics.

We will attempt to understand the rudimentary character of what one should
typically expects to see for such paths. We study
$N$ particles arranged randomly within some volume at time $t=0$. At some later time, $t=T$, we
require that the system returns to
a specified final configuration (which can even be the same as the initial one). We ask what kind of trajectories
are consistent with these boundary conditions at two time slices. For simplicity, we will consider
that there are no additional boundary conditions in space -- for example, there are no periodic or hard
wall boundary conditions, and the particles are free to move anywhere in d-dimensional space
$(-\infty,\infty)^d$.

\section{Model}
\label{sec-model}

As a concrete model, consider $N$ particles of equal mass, interacting via a two-body potential in $d$ dimensions; in this work
we've considered $d=2$ and $d=3$ %The initial position of the {\em mth} particle is $\br_{m,i}$ and the final is $\br_{m,f}$. 
with a number of different types of potentials. Two salient examples are, first, short ranged two body potentials of form 
\[
   \label{eq:Vpowerlaw}
   V_{power}(r) = C/r^p,
\]
where $p=3$ and $C$ is a constant; second we can consider ``soft" potentials that do not diverge at $r=0$:
\[
   \label{eq:Vsoft}
   V_{soft}(r) = C/(r_0^2+r^2)^{p/2},
\]
where $r_0$ is the length scale where the potential begins to flatten, and $p=3$. 

Because the final positions are given, there is no need to impose a limitation on particle
positions, so for example, periodic boundary conditions are not necessary. We rescale time
so that mass of each particle is $1$.

\section{Numerical work}
\label{sec-num}

We have been unable to find any significant literature on the solution of problems of this form, and it turns out to be quite challenging numerically. We have, however, devised a computational scheme (discussed in Appendix \ref{app:num_method}) to successfully find classical trajectories with these kinds of boundary
conditions for a small number of particles. 

Before examining these solutions is worth discussing the fact---obvious upon reflection---that setting state-to-state boundary conditions will lead to multiple solutions.  Imagine hard-sphere interactions between two particles.  Given fixed initial and final positions, the particles could either just move at constant velocity between their initial and final positions {\em or} they could bounce once.  For three particles, each could bounce with zero, one, or two others, and possibly even the same particle twice.  So the trajectories proliferate quickly with increasing particle number.  They are, however, finite in number and we expect this to be the case more generally~\cite{galperin1981,vaserstein1979systems,burago1998uniform,burago2000geometric}.

\begin{figure*}[htb]
\begin{center}
   (a) \includegraphics[width=0.4\hsize]{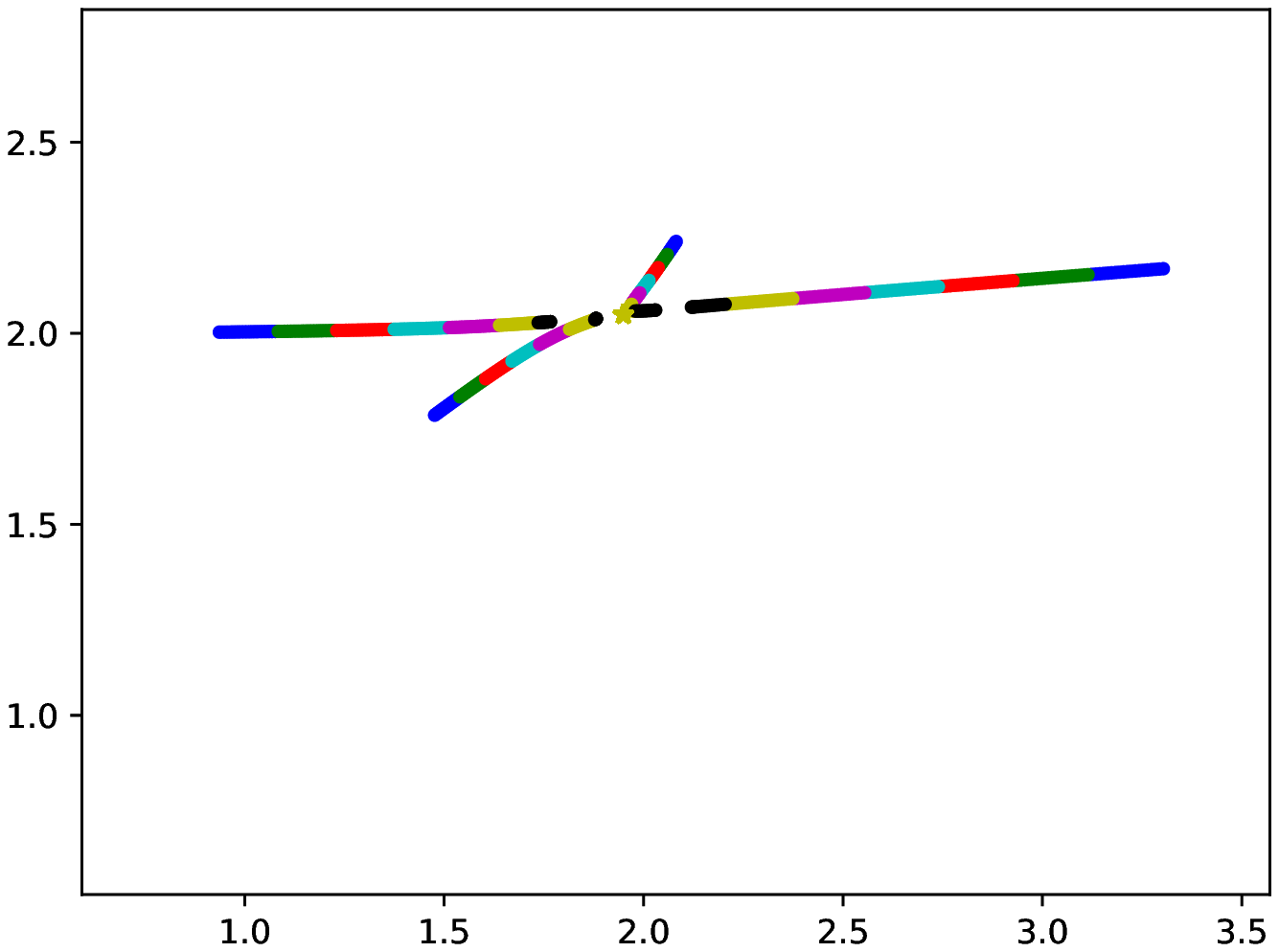}
   \includegraphics[width=0.4\hsize]{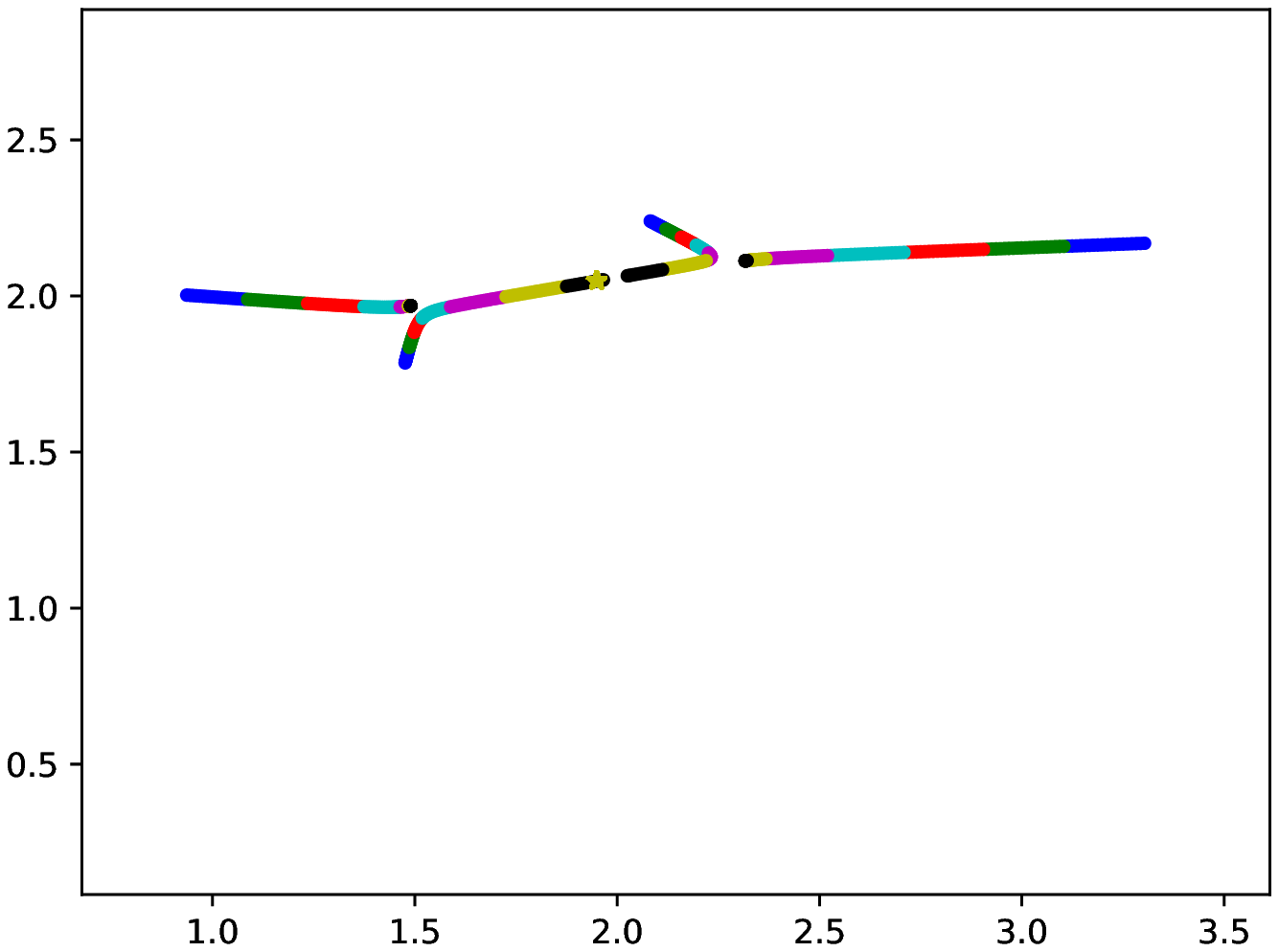}\\
   (b) \includegraphics[width=0.4\hsize]{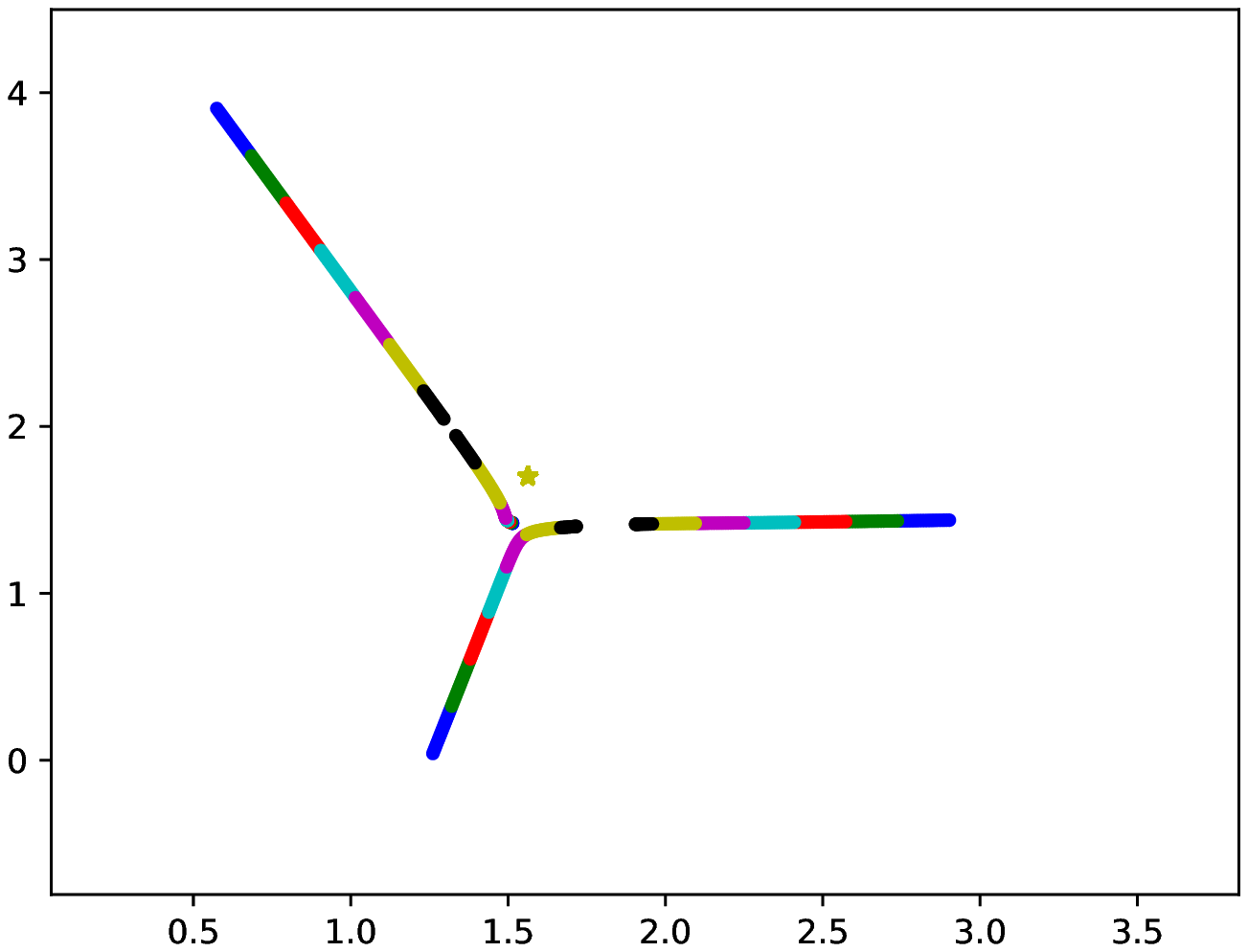}
   \includegraphics[width=0.4\hsize]{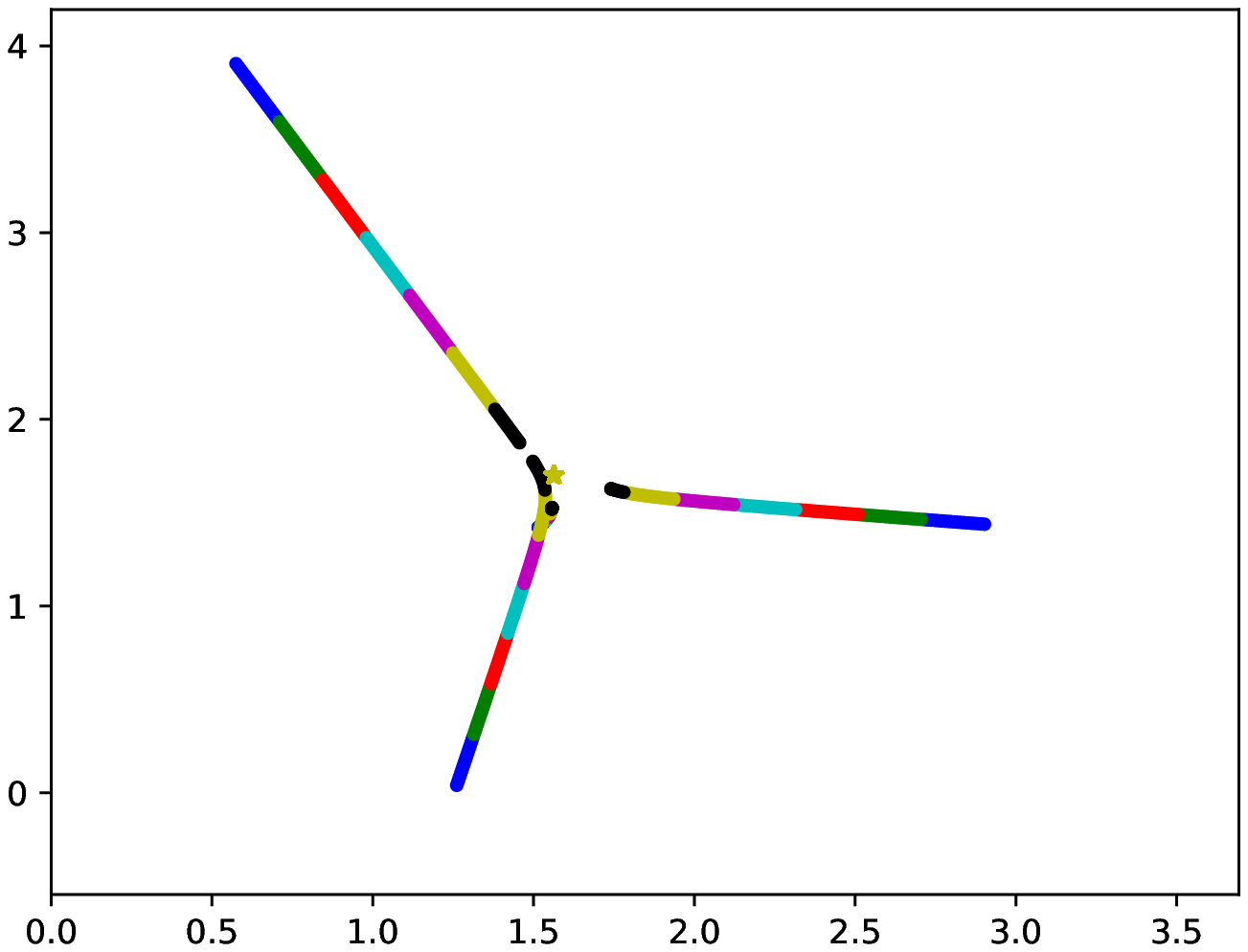}
\caption
{ 
(a) Trajectories for
four particles in three dimensions, interacting via two body repulsive power law potentials, 
Eq. \ref{eq:Vpowerlaw}, starting out in random initial positions with final positions that are be the
same as the initial positions. Each row shows trajectories for the same random initial condition. 
The path is symmetric in time, precisely doubling back on itself
after the half way mark. The different colors represent different time
windows. Blue is the earliest and black is closest to the halfway mark. 
The yellow star represents the trajectory
of the center of mass, which does not move, as expected. (b) The same parameters but with a different random initial condition.
}
\label{fig:4parts}
\end{center}
\end{figure*}

\begin{figure*}[htb]
\begin{center}
   (a) \includegraphics[width=0.4\hsize]{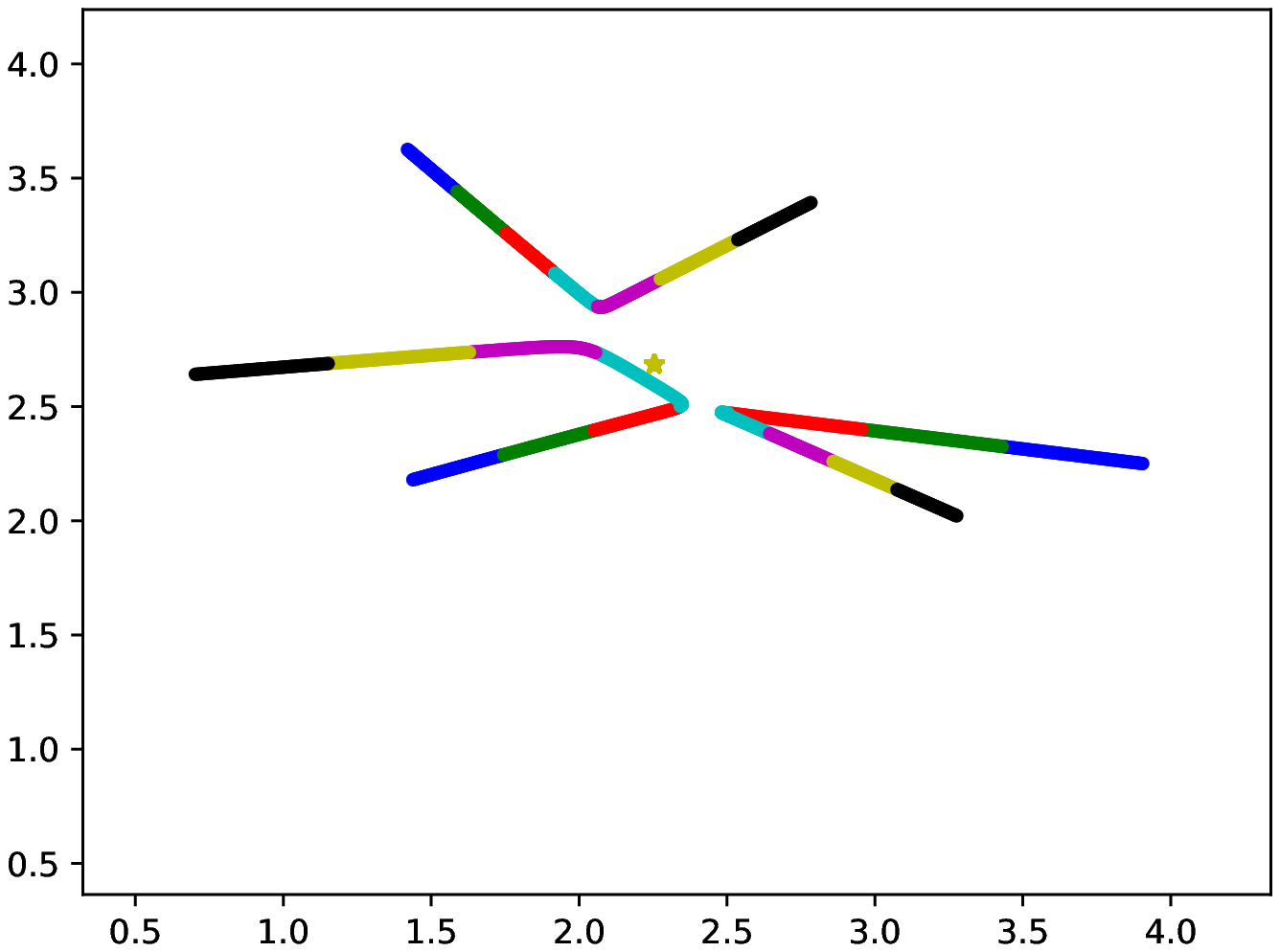}
   (b)  \includegraphics[width=0.4\hsize]{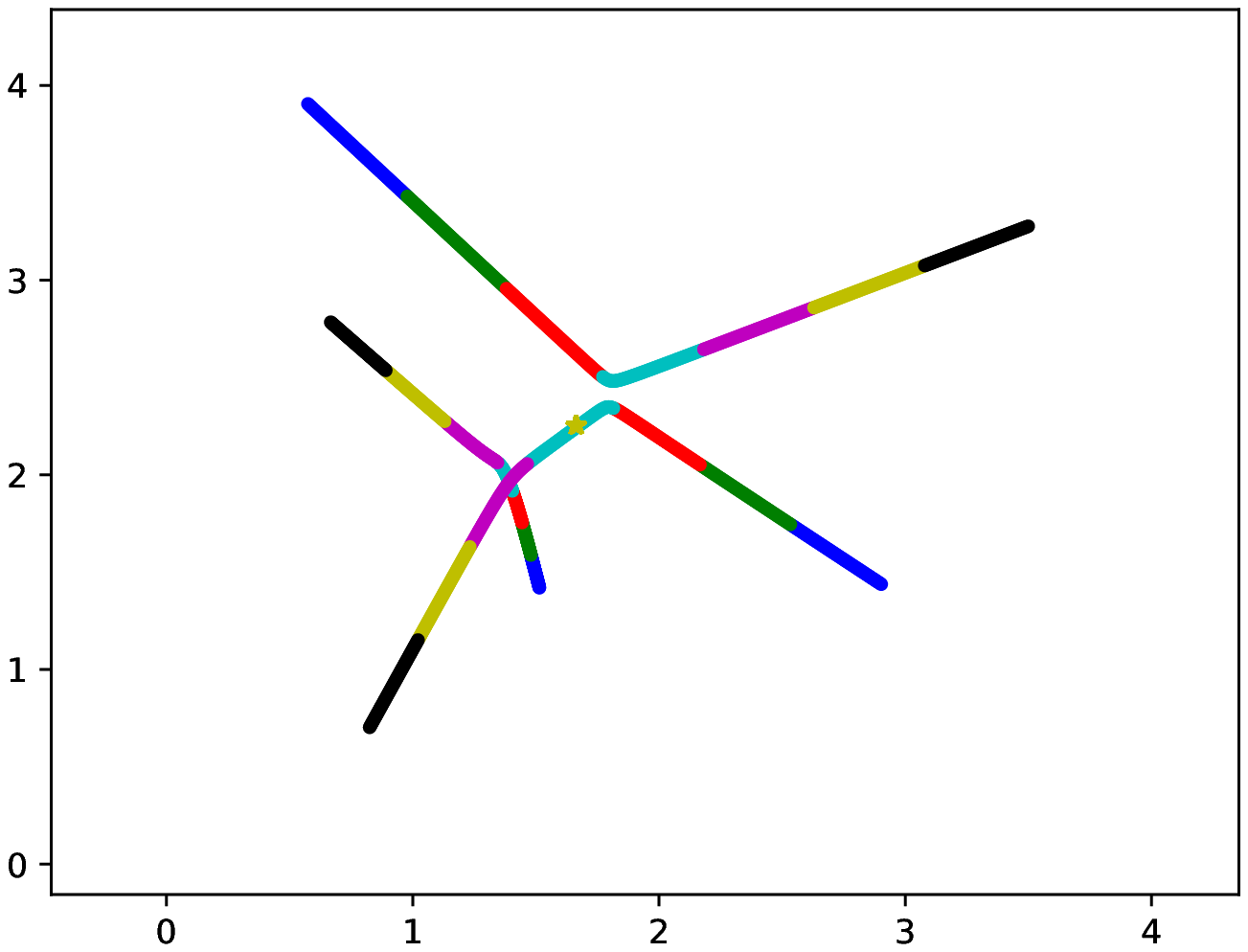}
\caption
{ 
(a) Trajectories for
three particles in three dimensions, interacting via two body repulsive power law potentials, 
Eq. \ref{eq:Vpowerlaw}, starting out in random initial positions with final positions that are not the
same as the initial positions. The different colors represent different time
windows. Blue is the earliest and black is closest to the end of the trajectory. The system is shown in the center of mass frame; the yellow star represents the trajectory
of the center of mass (b) The same system but viewed from a direction perpendicular to (a).
}
\label{fig:3parts_diff_ends}
\end{center}
\end{figure*}

\begin{figure*}[htb]
\begin{center}
   (a) \includegraphics[width=0.4\hsize]{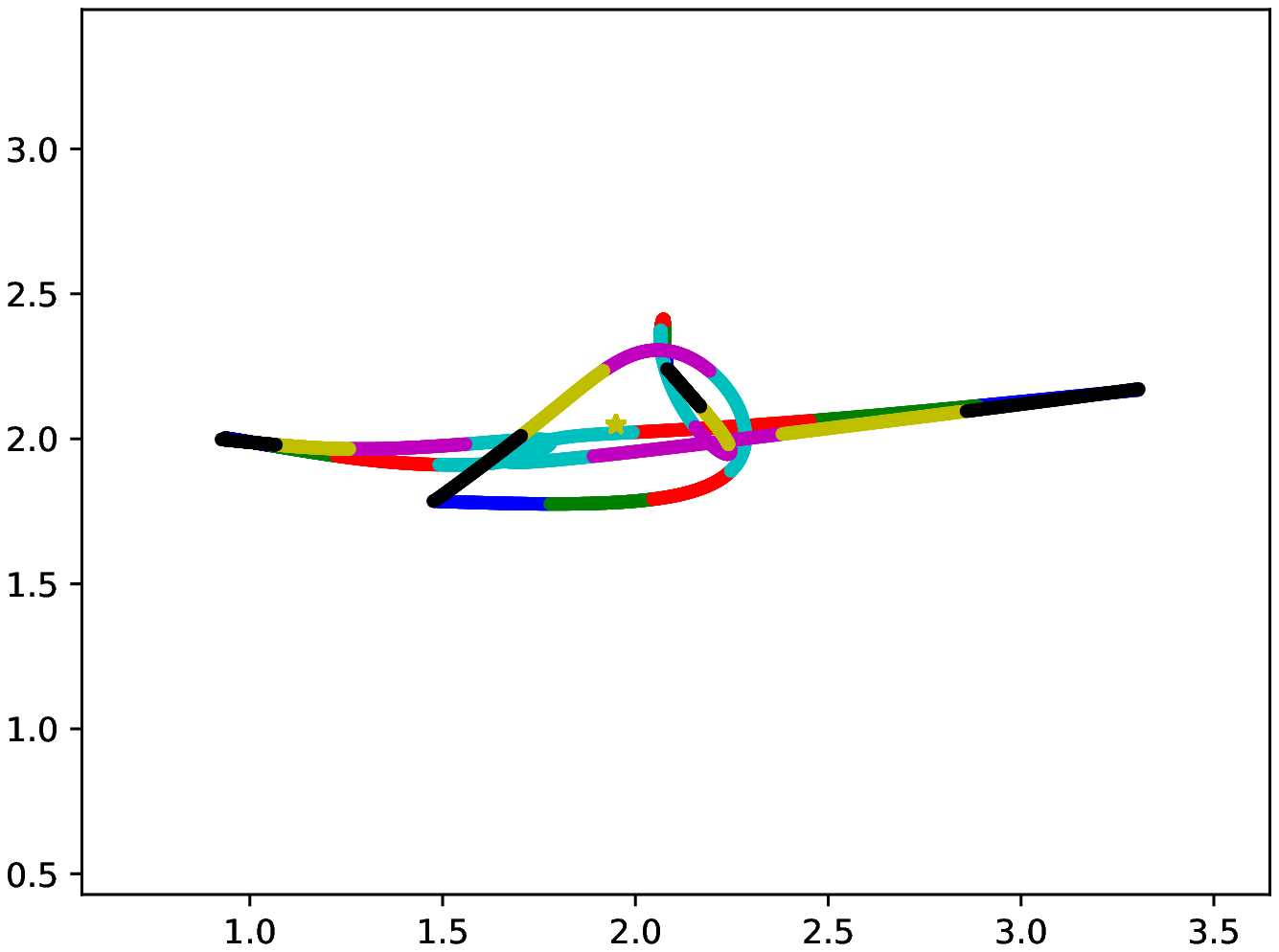}
   (b) \includegraphics[width=0.4\hsize]{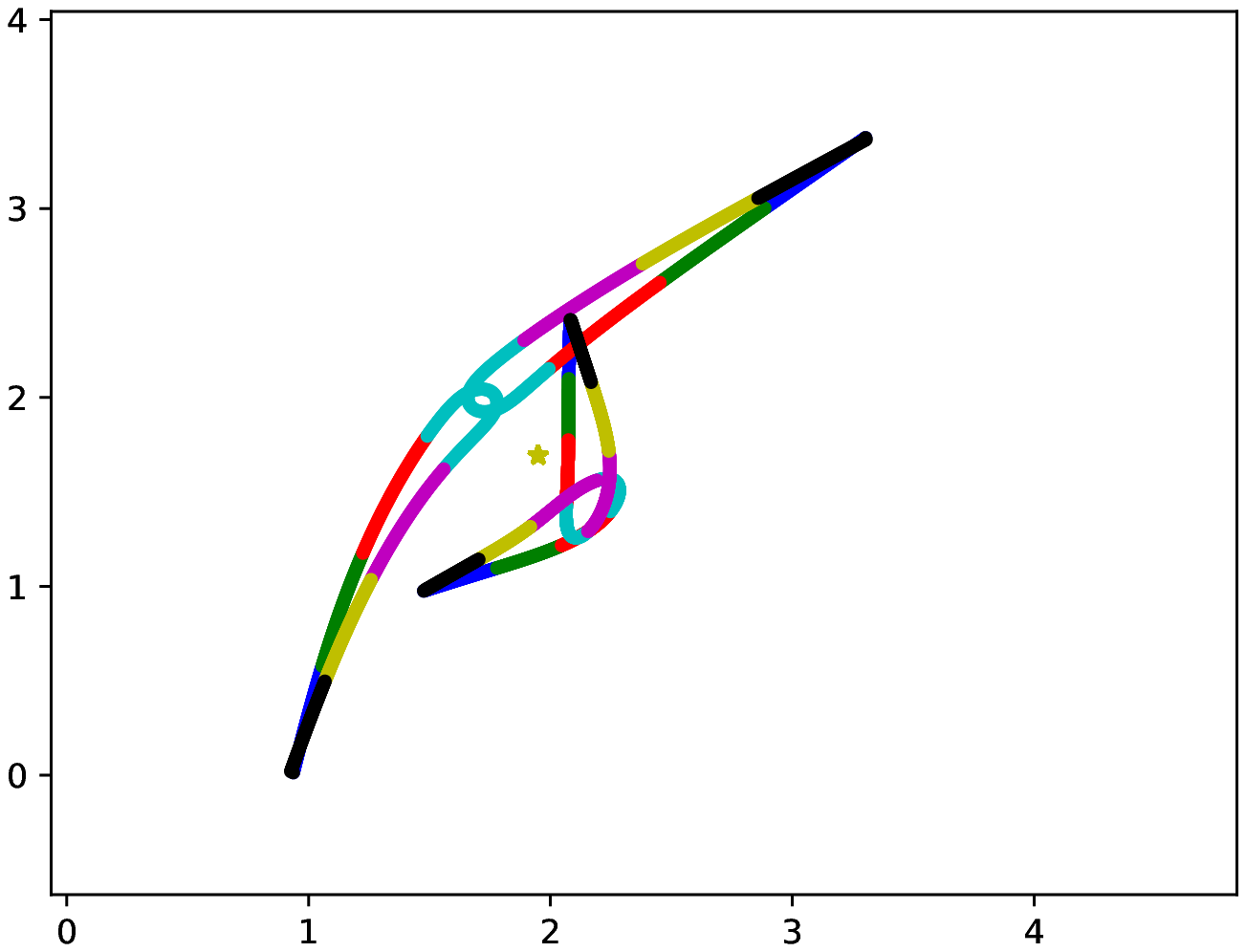}
\caption
{ 
Trajectories for
(a) Four particles in three dimensions, interacting via two body attractive  power law potentials, 
Eq. \ref{eq:Vpowerlaw}, starting out in random initial positions with final positions that are be the
same as the initial positions. The different colors represent different time
windows. Blue is the earliest and black is closest to the end of the trajectory. 
The yellow star represents the trajectory
of the center of mass, which does not move, as expected. (b) The same trajectory viewed at a 90 degree rotation.
}
\label{fig:4parts_attr}
\end{center}
\end{figure*}

We turn now to numerical solutions with a power-law inter-particle potential, as in Eq.~\ref{eq:Vpowerlaw}, with $p=3$.
Fig. \ref{fig:4parts} shows two possible paths (left and right panels) of four particles in three dimensions, where the initial and final
configurations were chosen randomly and are identical.  The trajectories precisely double back on themselves at the halfway
point. There are other trajectories possible as well---for example with the particles barely moving, staying in almost
exactly fixed positions. Figure ~\ref{fig:3parts_diff_ends} shows an example of a system where the same parameters as in Fig.~\ref{fig:4parts_attr} but with three particles and with different initial and final conditions.

In all of these trajectories the particles come together, interact, then move apart to ``meet" their final positions; we will argue below that this is very generic. To characterize compactness we can use the radius of gyration $R_g$, that is, the root mean square distance from the center of mass of a configuration. For the example in Fig.~\ref{fig:3parts_diff_ends}, the initial radius of gyration $R_g^2 = 1.081859$ and the minimum is $R_g^2 = 0.188861$.

We also examine the case
of attractive potentials in Fig.~\ref{fig:4parts_attr}. Here the trajectories are not time symmetric, but the particles
do come closer together for intermediate times. The radius of gyration of the system starts out at $R_g^2 = 0.847199$ and reduces
down to $R_g^2 = 0.077652$ before returning to its initial positions.

\begin{figure*}[htb]
\begin{center}
    (a) \includegraphics[width=0.4\hsize]{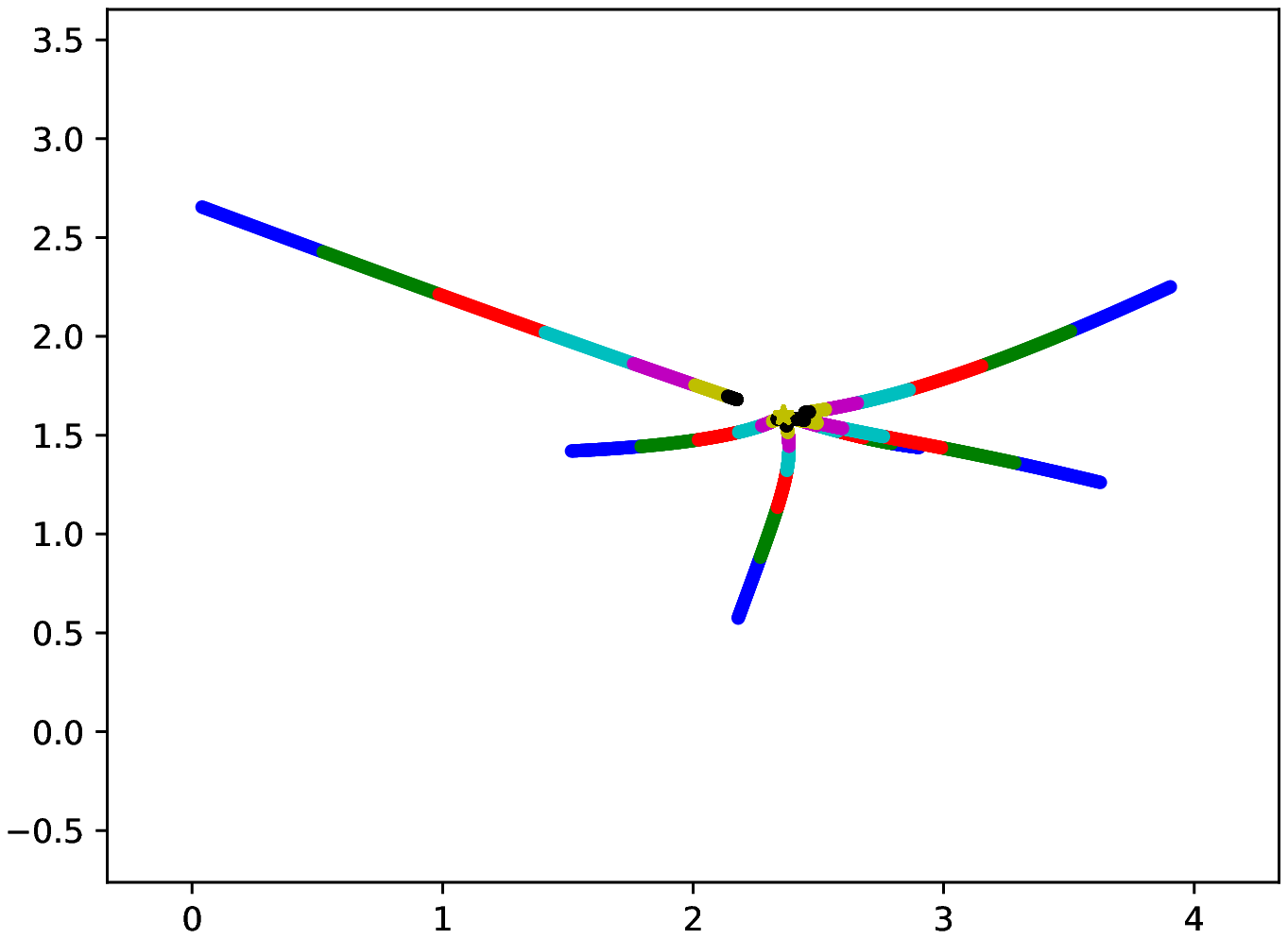}
   (b) \includegraphics[width=0.4\hsize]{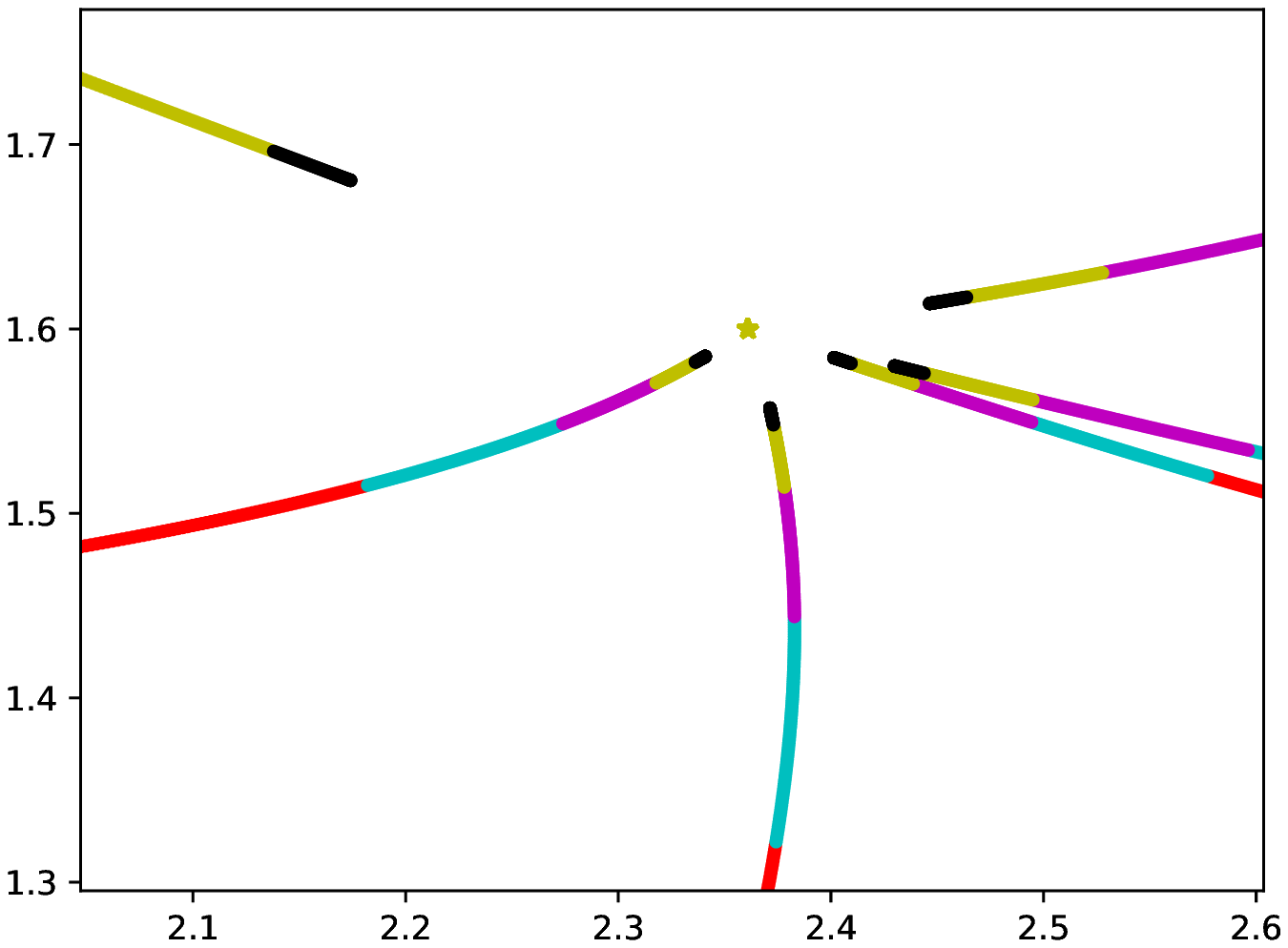}
\caption
{ 
(a) Trajectories are shown 
for
six particles starting out in random initial positions in two dimensions, with final positions that are the
same as the initial positions. The particles interact via two body soft potentials
given in Eq. \ref{eq:Vsoft}.
(b) The same trajectory enlarged  at the center. 
The yellow star represents the trajectory
of the center of mass.
As above, the path is symmetric in time, precisely doubling back on itself
after the half way mark.  The different colors represent different time
windows. Blue is the earliest and black is closest to the halfway mark. 
}
\label{fig:6parts}
\end{center}
\end{figure*}

Finally we consider a soft potential as in Eq. \ref{eq:Vsoft}. 
The two-dimensional trajectory shown in Fig. \ref{fig:6parts} has six particles. It demonstrates that there are parameters where particles come close together into high density regions. In the limit where the distance between the particles
becomes small compared with the range of the potential, this can be analyzed analytically; (see Sec.~\ref{sec:SoftPotentialDynamics} below.) After the
point of closest approach, the trajectory exhibits an affine expansion of particle positions with a time window during which expansion is exponential.

\section{Large N}
\label{sec-largeN}

Our method makes it computationally infeasible to determine numerical trajectories for large $N$, but we shall present a simple
theoretical analysis of this situation.
As discussed above the number of possible trajectories between a given initial and final configuration increases rapidly with $N$.  Here we argue that in a simple model (and probably more generically) the behavior of particles clumping together in a central region at intermediate times is generic, i.e. strongly dominates the set of trajectories. The basic argument, going back to hard spheres, is that more collisions leads to more ways to assemble a trajectory. The closer together particles are packed, the more collisions each one can undergo between the initial and final times, so highly-clustered trajectories should be most numerous. 

To ground this heuristic argument in more detail (and show further that most trajectories have the particles clustered in {\em one} big clump), we will consider a dilute gas of $N$ classical particles interacted via a short range two body
potential.  We assume that the particles start off in a region of size $L^d$ (where $d$ is the
dimension of space) in a typical random configuration and that after
a time $T$ they return to the same configuration.\footnote{This is no loss of generality; we can also generalize this analysis to
consider them returning to any other typical random configuration.} By ``typical" we mean that the density
of particles $n_L$ ($= N/L^d$ for the full box) is approximately homogeneous on a scale much larger than the mean inter-particle separation $L/N^{1/d}$, and no other special considerations are made in distributing the particles.

Because the interactions considered in this section are repulsive, and there is no spatial
confining boundary conditions, particles cannot form bound systems or become confined to a small region: they can, in principle, interact a finite number of times
before reaching their final configuration, but there are no forces preventing them from escaping to
infinity.  
Therefore we expect the average speed of particles in this situation is $v_{ave} \sim L/T$. 

To determine the cross section for a collision, that is, the effective range of interaction 
at which two incoming particles will have their
velocities noticeably perturbed, we consider the distance between two particles where the potential and kinetic energy become
equal.\footnote{This is called the ``Bjerrum length'' in plasma physics.} 
For example, for Eq. \ref{eq:Vpowerlaw} the (typical) radius $r_I$, of interaction is given by solving
\[
   \label{eq:V_R_I_1_2_mv2}
   V(r_I) = \frac{1}{2}m v_{ave}^2. 
\]
The cross section, which in three dimensions is an area, is then
\[
   \label{eq:def_sigma}
   \sigma \sim r_I^{d-1}.
\]

If the potential is soft, that is, it has a maximum value $V_{max}$, then there will be no real
solutions to Eq. \ref{eq:V_R_I_1_2_mv2} for sufficiently large $v_{ave}$. Thus there is a timescale
$T_{min} \sim L/\sqrt{2 V_{max}/m}$, below which particles are unable to travel distances of order
$L$ that are required for the analysis presented below to be valid; we shall assume that the time between the initial and final states exceeds this value.

Given this model, we can analyze the distribution of different types of solutions, which will indicate that the vast majority of trajectories collapse halfway, close to $t = T/2$.

\begin{figure}[htb]
\begin{center}
   (a) \includegraphics[width=0.4\hsize]{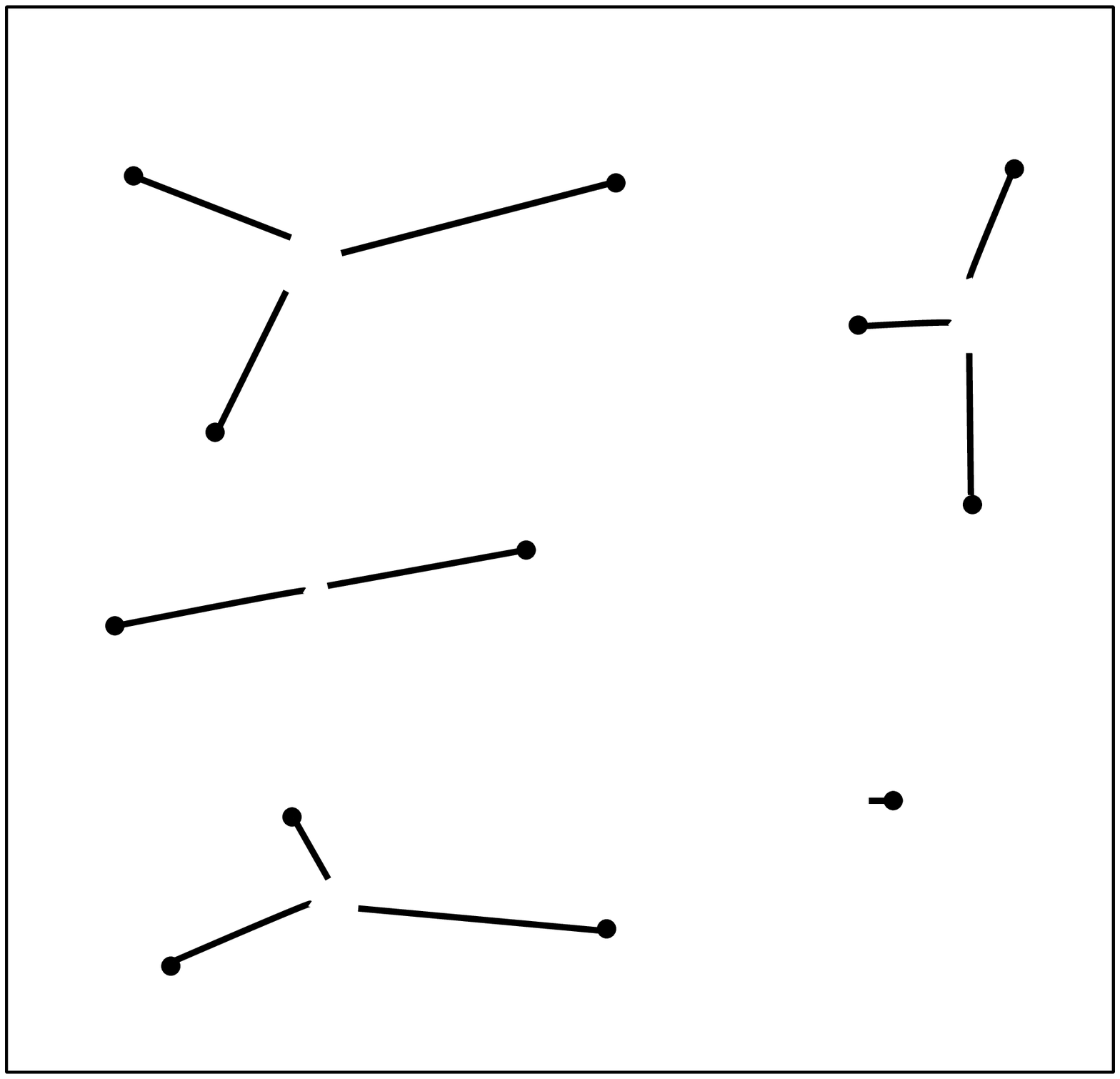}
   (b) \includegraphics[width=0.4\hsize]{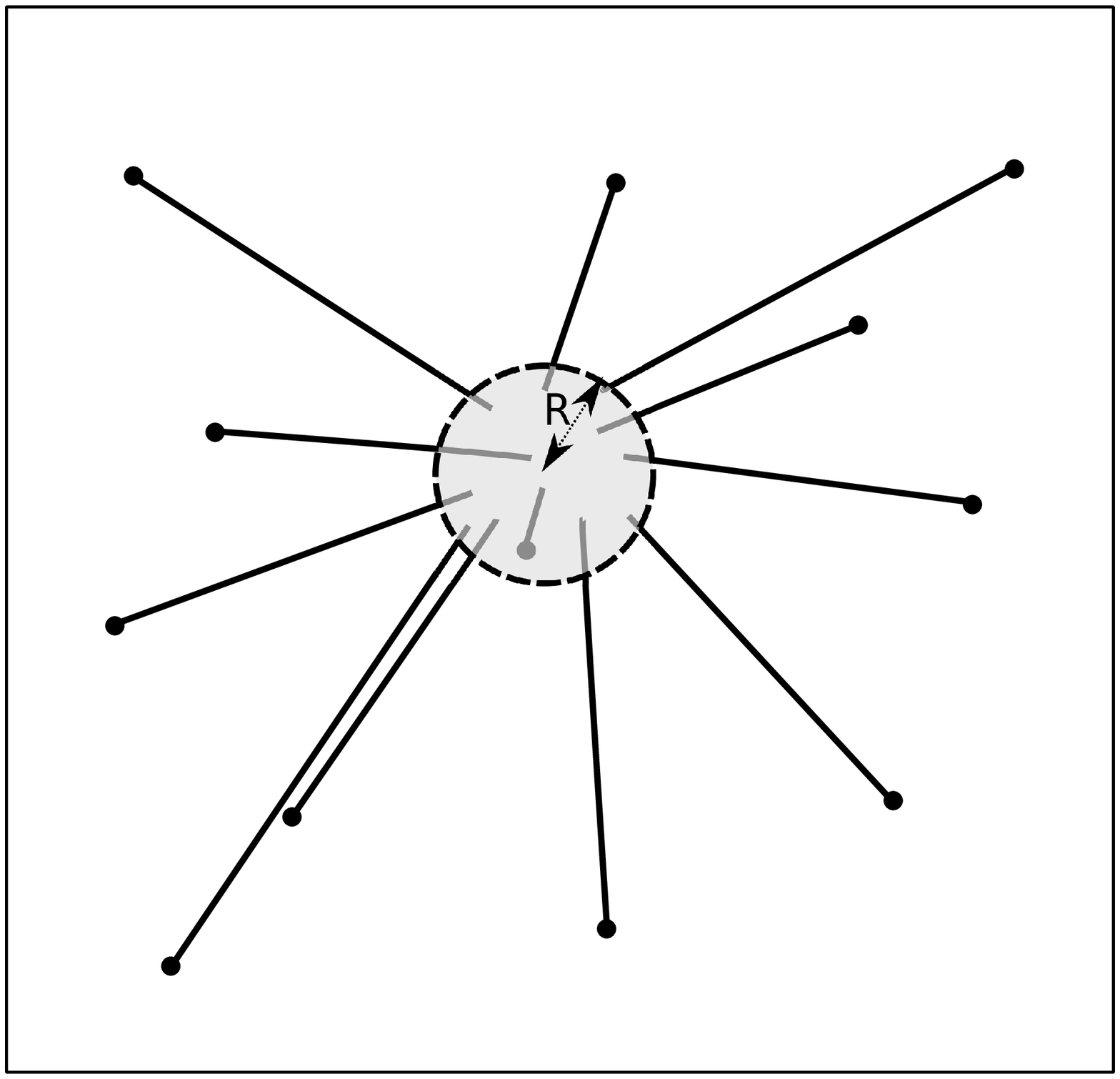}
\caption
{ 
   (a) {\bf Many-clump trajectories.} Time evolution of a system of particles with initial positions shown as black dots, move towards each other on the trajectories
shown. The collisions involve a small number of particles.
   (b) {\bf One-clump trajectories.} The time evolution involving a collapse of all particles in the system to a small region.
}
\label{fig:clumps}
\end{center}
\end{figure}

We start by considering two different kinds of system trajectories, as shown in Figs. \ref{fig:clumps} (a) and (b).
A large class of system trajectories will typically look something like what is shown in Fig. \ref{fig:clumps}(a), that 
we call ``many-clump" trajectories.
They can be thought of as being comprised of many separate collisions of a relatively small number
of particles. These are the sort of trajectories that would dominate if the particle distribution stayed diffuse between the initial and final states.
This should be contrasted with trajectories of the ``one-clump" type, illustrated in Fig. \ref{fig:clumps}(b). 

In brief, in we will first estimate the number of trajectories of a one-clump collision by estimating the number of collisions undergone by particles as a clump doubles (or halves) in size. Then, for many-clump systems, we will estimate combinatorially the number of distinct ways particles can be grouped into clumps.  Combining this with the previous result we estimate the total number of many-clump trajectories and show that there are still less than one-clump trajectories, 
so that in the large-$N$ limit one-clump trajectories dominate.

\subsection{Counting trajectories in a cluster}
\label{sec:ClusterTrajectories}

Here we ask how many trajectories 
there are if the system collapses into a small roughly spherical region of radius $R$, as shown in
Fig. \ref{fig:clumps}(b).
Here
the particles travel into a smaller region of minimum radius $R$, at times close to $t= T/2$, after
which they expand out.  We first estimate
how many collisions they undergo during this process.

We can ask how many collisions a single particle will undergo in the time it takes the sphere to expand to double its size, from
radius $R$ to $2 R$, i.e. while the system size is $\sim R$. The number density is $n_R \sim N/R^d$, and a particle travelling in a straight line at
velocity $v_{ave}$ covers a distance $\sim R$. 
The number of collisions per particle $N_c$ is then $n_R$ times volume of
a cylinder of length $R$ and cross section $\sigma$, or
\[
   \label{eq:Nc_sim_NR1-d}
   N_c \sim (N/R^d) R \sigma \sim N R^{1-d} \sigma.
\]

The total number of collisions increases with decreasing $R$ until a point where the gas is no longer dilute and the
inter-particle spacing $R/N^{1/d}$ is the distance of closest approach $\sim r_I \sim \sigma^{1/(d-1)}$. 
Therefore, for a potential that is divergent at small distances (so that $r_I$ is
meaningful), $R_{min} \sim r_I N^{1/d}$.  Thus the number of collisions per particle while
they are in a maximally compact configuration
is 
\[
   \label{eq:N_c_sim_N_1overd}
   N_c^{max} \sim N^{1/d}.
\]
This means that the total number of collisions suffered by a particle when inside a ball of radius $2R_{min}$ or
less increases in three dimensions as $N^{1/3}$. This makes sense, because this corresponds
to the width of a cube closed packed with $N$ particles. A given particle is able to collide with effectively a single straight line of particles during that time. The number of collisions rapidly falls off as $R$ increases as given in Eq.~\ref{eq:Nc_sim_NR1-d}, so the total collision number is dominated by this close-packed period.

Having an estimate for the total number of collisions, we estimate the number $\mathcal{N}_O$ of independent system
trajectories of the one-clump type. We do so by estimating a similar quantity, the number $N_{PO}$ of periodic orbits (PO) of period $t$ on a chaotic attractor.  This is related to the ``topological entropy'' $H_{Top}$ and the Kolmogorov-Sinai entropy $H_{KS}$.

There are a number of rigorous results relating $H_{Top}$ to $N_{PO}$, for discrete time
dynamical systems~\cite{bowen1970topological,katok1980lyapunov}.
Similar results have been proved for diffeomorphisms, two dimensional Sinai Billiard maps~\cite{baladi2020measure,buzzi2018degree} and two
dimensional hyperbolic billiards~\cite{chernov1991topological}. In the limit of large time $t$, or discrete time iterations $n$,
\[
   N_{PO} \ge \exp(n H_{Top}).
\]
In some systems, this has been shown to be an equality~\cite{bowen1970topological}.
We will assume that this result is true even in the case of hard spheres in $d=3$, although this apparently has not been proven.

It is also generally the case that the topological entropy is greater than or equal to the Kolmogorov Sinai entropy~\cite{chernov2000entropy}, and 
often they are equal~\cite{baladi2020measure}.  In term of continuous time dynamics, then,
\[
   N_{PO} \ge \exp(t H_{KS}).
\]
Below we will conservatively assume that this is an equality, because in many cases it is, and even if not it would then give an underestimate of trajectories of this type.

Now, this result applies to periodic (unstable) orbits of Sinai billiards; this is not identical to the problem considered here, because the return to the same point in configuration space
is not the same as a periodic orbit. However the underlying reason for this proportionality is the
mechanism of kneading (folding and stretching) in phase space. Every new fold creates another set of layers and each layer creates a new fixed point that is a periodic orbit. The number of layers increases exponentially in time and is measured by $H_{KS}$.
Therefore we estimate the number of independent system trajectories the same way~\footnote{
This leaves off a factor of $f_R$ giving 
the fraction of periodic orbits that at some point contract to radius $R$. $\log f_R$
is expected to scale as the entropy difference in going from the initial density $n_L$ to the final
density, and should therefore be extensive, that is proportional to $N$. We will see below that this
is subdominant and will not be considered further.
}:
\[
   \label{eq:N_Rsim_exp_tH}
   \mathcal{N}_O \sim N_{PO} \propto \exp(t H_{KS}).
\]
This extension to trajectories that have fixed rather than periodic boundary conditions appears plausible, but has not been proven.

The Kolmogorov Sinai entropy has been studied in dilute hard spheres~\cite{de2005kolmogorov}.
To leading order in the density $n$,
\[
   H_{KS} \propto N \nu [-\log (n a^d) + B + O(n a^d)],
\]
where $a$ is the the particle diameter (which in this context is $2 r_I$ with $r_I$ being the particle radius), $\nu$ is the single particle collision
frequency, and $B$ is a constant. We are interested in $H_{KS}$ when the particles become
close to each other, in which case $n \sim 1/r_I^d$, so all three terms are roughly $n-independent$ so omitting numerical pre-factors we have
\[
   H_{KS} \propto N \nu.
\]

We can approximate $\nu$ as the ratio of the the total number of collisions suffered by a single particle to the time during which the system is inside a ball of radius $2R$: $\nu = N_c/t_R$.
From above, $\exp(t H_{KS})$ is an estimate of the number of independent system trajectories, and we would like to find the $R$ where this is maximal. This requires maximal $H_{KS}$ and in turn maximal $N_c$.
We saw from Eq.~\ref{eq:Nc_sim_NR1-d} that this requires minimizing $R$, in which case we see from Eq. \ref{eq:N_c_sim_N_1overd}
that $N_c^{max} \sim N^{1/d}$.

Therefore from Eq. \ref{eq:N_Rsim_exp_tH} we see that the total number of one-clump trajectories is
\[
   \label{eq:N_b_sim_exp}
   \mathcal{N}_O \sim \exp[C N^{1+1/d}]
\]
where $C$ is a constant. 

We now compare this to the {\em many}-clump scenario, where each clump is governed by Eq.~\ref{eq:N_b_sim_exp}, but in which there are many way to arrange the interactions into a spectrum of clumps.

\subsection{Cluster combinatorics}
\label{sec:combinatorics}

For the many-clump trajectories depicted in Fig. \ref{fig:clumps}(a), we can classify the trajectory of the complete system by the particles 
that collide with each other.  In this case, particles can be divided up into clusters of interactions. Suppose there are $M$ clusters in total
and $m_i$ particles that collide in the $ith$ cluster. Then the total number of arrangements of
clusters is
\[
   \label{eq:groupN_M}
   \frac{N!}{\prod_{i=1}^M m_i! \prod_{j=1}^N n_j!}.
\]
Here $n_j$ is the total number of clusters that have size $j$. 

In fact this is an overestimate, because it assumes that particles at opposite ends of the box
can collide, whereas they will typically  be excluded from doing this by other particles that are in their
way. This tends to happen if their separation exceeds the  
$l = 1/(\sigma n_L)$.
Because of this bound in the range of the interaction, the number of ways of grouping
collisions is the same as a gelation or percolation problem. Every particle has the choice to group
with any particle within a mean free path; these number $N_l \equiv n_L l^d \sim n_L^{1-d} \sigma^{-d}$. The number of ways of
arranging particles inside a volume of order $l^d$ is similar to Eq. \ref{eq:groupN_M}, and should
be bounded by $N_l!$, therefore the entropy of this region is bounded by $N_l\log N_l$. This gives an
entropy per particle of $\log N_l$, or a total gel entropy of $S_G \sim N \log N_l$. Therefore the
number of independent cluster configurations is 
\[
   \label{eq:NC_sim_expNlogNL}
   \mathcal{N}_C \sim \exp(N\log N_l) .
\]

Now we are in a position to combine these combinatorics
with the number of trajectories in a single clump (Eq.~\ref{eq:N_b_sim_exp}), to analyze more fully multi-clump trajectories.

\subsection{Combining cluster combinatorics and trajectory number}
\label{sec:ClusterAndTrajectories}

We can already see that the supra-exponential behavior of one-clump trajectories, Eq. \ref{eq:N_b_sim_exp}, will dominate over
the number of multiclump configurations, Eq. \ref{eq:NC_sim_expNlogNL}, favoring one-clump trajectories. We will now analyze what
happens for intermediate clump sizes, where the effects of the combinatorial multiplicity and
trajectory counting of clumps are both accounted for. As described above, a particle that is part
of one clump cannot travel for more than the order of the mean free path $l$, through another clump.
But we can make large clumps of dimensions greater than $l$, by having all particles in a clump pointing directly towards a central region.

For a single system trajectory with clumps of size $M$, there will be $\sim N/M$
separate clumps. 
For each of these clumps, the number of separate collision trajectories, from  Eq. \ref{eq:N_b_sim_exp},
is then $\sim \exp(C M^{1+1/d})$

Therefore the number of distinct trajectories for all of the clumps is
\[
   \mathcal{N}_T \sim [\exp(C M^{1+1/d})]^{N/M} = \exp[C N M^{1/d}]
\]
for a single configuration of clumps.

But now we need to multiply this by the number of
possible configurations when clumps are of
size $M$, to find the total number of trajectories (after which we can check for which $M$ this is maximized.)
Let us call this total number of clump configurations $\mathcal{N}_C(M)$. We will give a calculation of this in
Appendix \ref{app:calc_fm}, but its detailed form is is not necessary in order for us 
to draw our final conclusion concerning the dominance of
one-clump trajectories---general properties of $\mathcal{N}_C(M)$ will suffice.
We see from Eq. \ref{eq:NC_sim_expNlogNL} that with no restriction on the size of clumps, 
the entropy $S_C \equiv \log \mathcal{N_C} = N \log N_l.$ Any restriction on clump size will only
decrease this, so that $S_C(M) \equiv  \log \mathcal{N_C} (M) = N f(M).$ This follows from the fact
that we expect that as we increase $N$ with fixed clump size $M$, the entropy will
be extensive in $N$. Also in the limit where $M=N$, there is only one clump, so $f(N) = 0$. Therefore we
expect $f(M)$ to be a decreasing function of $M$.

We can now put this together: for clusters of size $M$, the total number of system trajectories
including all arrangements of clumps of size $M$, is
\[
   \mathbb{N}_{M} = \mathcal{N_C}(M) \exp(C N M^{1/d})  = \exp(N (f(M) + C M^{1/d}))
\]
Because $f(M)$ is a decreasing function of $M$ bounded by $\log N_l$ (see Eq. \ref{eq:NC_sim_expNlogNL}),  $\mathbb{N}_{M}$
is maximized when  $M = N$.
This means that the overwhelming number of trajectories will be in the form of a single large clump, as in Fig. \ref{fig:clumps}(b).

Therefore as the number of particles in the system $N\rightarrow \infty$, if trajectories are given equal weight then trajectories
of the one-clump type, shown in Fig. \ref{fig:clumps}(b), will be seen with probability approaching $1$. In the case of repulsive interactions with a finite separation, the minimum radius $R$ of this system
approaches the scale of close packing.

\subsection{Soft potential dynamics}
\label{sec:SoftPotentialDynamics}

For soft potentials where the particles can become arbitrarily close (
for example the potential defined in Eq. \ref{eq:Vsoft}) we can say a bit more. Consider the limit where all of the
particles are close enough to each other that the potential is well described by a quadratic
\[
   V_q(r) = -\frac{k}{2} r^2 + V_{soft}(0),
\]
where $k = -V_q''(0)$. Particles now interact via unstable forces linear in their vector separation, i.e. the
force on the $ith$ particles is
\[
   \label{eq:LinearForces}
   \bF_i = k \sum_{j=1}^{N} (\bx_i-\bx_j)
\]
where we are using bold to denote vector quantities. Because there are no external
forces acting on the system, momentum is conserved, so we can choose a reference frame where the
center of mass of the system is zero. In this case, Eq. \ref{eq:LinearForces} together with
Newton's second law becomes
\[
   m \ddot{\bx_i} = N k \bx_i.
\]
Each particle evolves independently under the influence of an unstable linear force field
so that solutions are linear and of the form $\exp(\pm\gamma t)$, with $\gamma = N k/m$.
Particles close to each other expand out affinely going from an initial quadratic 
expansion to an exponential one. This exponential regime will cease to be
a good approximation when nonlinear terms in the force become important.

\section{Discussion}
\label{sec-discuss}

\subsection{Summary}

We have analyzed the problem of $N$ classical particles placed randomly in some region of size
$L^d$, which interact via short range potentials with no confining potential or
boundary conditions applied. We have examined primarily repulsive interactions.\footnote{ We do this because much
is known about the Kolmogorov-Sinai entropy of hard sphere systems, which are expected to behave similarly
to a large class of short range repulsive potentials. Not as much is known about attractive
potentials.} 

Instead of considering initial conditions where position
and momentum are specified at one time, $t=0$, we instead only specify positions, but
not the momenta. To make up for the lack of momentum specifications, we specify
the final positions at some later time $t=T$. We can make the initial and final positions
distinct but random, or make them the same. In either case, there are a finite number
of distinct system trajectories that satisfy these conditions. We have analyzed what
happens when $N$ is large and the initial and final configurations are that of a dilute
gas. 

We find that a system trajectory picked randomly from
all of the possible trajectories will be overwhelmingly likely to undergo a collapse at
intermediate times, reaching a density similar to close packing when 
the potentials are hard. If they are soft, it is expected that they will collapse even
further (although precisely how much they will typically collapse remains to be studied.)
In this regime, particles contract and expand affinely, obeying a linear differential equation 
with solutions that are exponential in time.

What was just described is very different than what one might naively have expected. Our
experience with statistical mechanics would instead suggest the configurations would
remain extended, to keep the entropy large, rather than collapsing to a much smaller
entropy, in apparent violation of the second law of thermodynamics. However, the situation here is quite different. In order to return to the specified (but generic) final positions, the initial momenta must be special rather than generic. The preponderance of trajectories that appear entropy-decreasing is an effect of these special momenta.

The different choice of boundary conditions allows for some level of indeterminacy 
because there are a vast number of trajectories consistent with these kind of boundary conditions
for large $N$. It is somewhat unclear how to interpret this indeterminacy or what ``chooses'' the trajectory; but if we accord equal weight to them we can turn relative frequencies directly into probabilities. This use of typicality is distinct but related to arguments
that are used elsewhere in statistical mechanics.

We have focused on short-range and repulsive interactions with no spatial boundary conditions in place.
The important qualitative feature
to the argument presented here is to have a lot of collisions when the particles form a high density cluster.
Collisions exponentially increase the folding of phase space, causing a proliferation
in the number of trajectories. These qualitative features are likely to survive
even if there is an attractive component to the potential, in addition to the hard core, and possibly even if the forces are long-range.  

In terms of boundary conditions, if boundary conditions were imposed that are sufficiently far from the $L^d$ box, there are still time windows and/or parameters where we can expect similar behavior.

Extension to more general interactions and scenarios would be an interesting direction for future work. 

\subsection{Field theory, gravity, and quantum mechanics}

Our analysis has applied strictly to particle mechanics with short-range interactions, and there are significant questions remaining even in that analysis.  Nonetheless in advance of further research it is interesting to speculate about a similarly-posed problem including quantum mechanics, field theory, or gravity.

\subsubsection{Quantum mechanics}

Because non-relativistic quantum evolution is first order in time,
it is usual to describe evolution by initial conditions at one time. However in the formalism of
Greens function propagators, initial and final positions at different times are given. Thinking
of this problem from the point of view of propagators, and its path integral formulation, leads
to the consideration of a semi-classical description based on expansion over classical paths.
The implication of the work described here is that the majority of classical paths take
trajectories that become strongly localized for some intermediate times. To compute the
semi-classical many-particle propagator for this requires evaluation of the phases and weights that all of these
paths contribute to the propagator. If indeed their collective effect is substantial, this could
have interesting implications for experiments.
If we consider a physical situation, such as a dilute gas of cold atoms interacting via short range
potentials confined to some $L^d$ region, if we switch off confining potentials, we could in
principle measure the statistics of particles remaining in that region after some time $T$. It may
be that the paths contributing to this situation are well described in terms of these one-clump
paths.

We also note that there is an interesting approach to the measurement problem in quantum mechanics that
involves having initial and final boundary conditions to a quantum system, the ``Two time interpretation of quantum mechanics"~\cite{aharonov2005two}. 
However this line of work, along with others studying quantum retrocausality~\cite{sep-qm-retrocausality}, are asking questions that appear separate from the ones considered here.

\subsubsection{Classical field theory}

If we consider some classical field, for example the evolution of a scalar field $\phi$
subject to a local potential $V(\phi)$ with quadratic coupling,
we might in general expect similar to behavior
to what we found for classical particles. In the case of classical particles,
if they are initially localized to a small region with a random distribution of velocities,
they expand out, occupying a larger portion of space as time increases. Correspondingly,
a random field configuration localized to a small region would expand in a similar way.

Now consider an analogous situation to the two time boundary conditions considered here. We start in some random initial field configuration
in a large box of size $L^d$ and then
apply a boundary condition at a later time $T$ requiring the field return to the same configuration,
or one that is statistically equivalent.
Preventing the field spreading outside of the $L^d$ box at the maximum propagation
velocity of the field would require special conditions on field derivatives.
This would cause the region of nonzero field to contract, in essentially the time-reversed dynamics
of the expansion just mentioned. The contraction would continue until nonlinear interactions
cause sufficient scattering for the waves to reverse their course, enabling
the return to the prescribed final state. We do not know to what extent this scenario would lead to a
conclusion similar to the case of classical particles. For example, are most trajectories of a
similar nature, or are they more spread out in this case? If indeed the majority of trajectories
would be of a strongly localized nature, we would expect $V(\phi)$ to have relatively large values in such
configurations, when compared to the values at times when the configurations were more extended.\footnote{If we extrapolate to the context of inflationary cosmology, this could give an explanation for why the field starts ``up the hill''
}

\subsubsection{Gravity}

Adding Newtonian or relativistic gravity into the mix adds additional subtleties.  As mentioned above, long-range attractive forces may act quite differently than the short-range repulsive ones assumed in Sec.~\ref{sec-largeN}, and it is unclear whether the basic argument still holds.  Moreover, the statistical mechanics of self-gravitating systems are less agreed-upon than for short-range forces, there in general being no equilibrium state; for example a roughly uniform distribution in particle positions is generic in both configuration space and phase space for a ``gas'' of particles with short-range interactions, but under gravity (for example) a ``generic'' configuration in phase space is very clumpy in position space and it is less clear that choosing the initial and final configurations to be roughly uniform is a non-fine-tuned specification of boundary conditions.

Still more troubling, General Relativistic systems with reasonable energy conditions are subject to singularity theorems that can forbid the sort of ``bounce'' from a ``collapsing'' to an ``expanding system'' contemplated here.\footnote{``Collapsing'' and ``expanding'' are relative to a direction of time, and as discussed below in this case time's arrow may reverse at the point of greatest compression so that both halves would in fact be considered ``expanding.'' This matters for some but not all relevant singularity theorems.}

Nonetheless, it seems plausible that {\em if} our result applies to fields and self-gravitating particles, it could underlie just the sort of ``time reversal'' inflationary cosmology proposed in~\cite{aguirre2003inflation}.  There, it was imagined as a low-entropy boundary condition on which the inflaton field is specified to be uniform but at large $V(\phi)$, then evolves to lower field values (while driving inflation) in {\em both} time directions, with the surface on which the field is specified then representing a reversal of the arrow of time.  By construction this scenario is nonsingular. (See~\cite{aguirre2003inflation} for how the singularity theorems are either inapplicable or evaded.) In the present context, we could imagine a late-time post-inflationary universe that is ``generic" at two times, in-between which most trajectories would pass through a phase in which the universe has small scale factor with $V(\phi)$ large.

\subsection{Implications for time's arrow}

Our everyday experience of the world as well as laboratory experiments correspond with a paradigm in which the world is in some state at a given time, and its future evolution is predicted using that state (in phase space). But this view makes it virtually impossible to reconcile (a) the second thermodynamic law with (b) non-fine-tuned boundary conditions for the universe.  That is, given the second law, the early state of the universe must be low entropy, and a state cannot be both special (low entropy) and generic (non-fine-tuned) in the same way.

But Cauchy boundary conditions are {\em not} those we consider in formalisms such as the action principle in classical mechanics or path integration in quantum mechanics -- in these one instead fixes an initial and final configuration. It is then worth entertaining the idea that this may also be appropriate for posing {\em cosmological} boundary conditions.  In this case, the paradox disappears: a generic trajectory for a generic set of boundary conditions for the universe passes through a very low entropy state.\footnote{Here we are considering definitions of entropy with positional
coarse-graining~\cite{vsafranek2019quantum}.} which we can take to be ``initial.''

It is worth emphasizing that this picture is not subject to the Boltzmann brain paradox.  That paradox would apply if we were to specify many different actually-existing ``initial'' high-entropy boundary conditions and look for the ultra-rare cases in which entropy decreases to some substantially lower minimum value before rising again. In this case, trajectories containing a low-entropy ``middle" configuration are generic rather than special, so brains will tend to find themselves having been born the natural way, evolved from a low-entropy past.

Both everyday experience and scientific experiment underpin a view in which there exists an objective past, but an uncertain future.  This has led to a scientific method based on predicting (because this is all we can do) the future from the present, whilst the past might be retrodicted but is also accessible by records.
This time asymmetry is widely believed to be closely tied to the second law of thermodynamics, which is in turn tied to cosmological boundary conditions.  However, in discussing those boundary themselves conditions we need not use the same paradigm as for other physical systems.  Alongside the ``Cauchy'' paradigm of physics has always been the parallel mode of the action principle and the path integral, which entail quite differently-posed boundary conditions. The main purpose of this work is to point out that this
different point of view, when actually applied even to simple mechanics systems, has surprising consequences.  We hope that
this will stimulate more work in the same direction.

\acknowledgements

We thank Richard Montgomery and Carles Simo for useful discussions. This work was supported by the Foundational Questions Institute \url{<http://fqxi.org>}.

\appendix

\section{Numerical Method}
\label{app:num_method}

We consider $N$ particles subject to a force field $F$, and want to solve
\[
   \label{eq:NewtonsLaw}
   m \frac{d^2 x_i}{dt^2} = F_i(\{x_j\})
\]
where $m$ is the mass, which we will take to be unity for notational simplicity. $i = 1,2\dots\,N d$
where $N$ is the number of particles and $d$ is the spatial dimension.
We impose boundary conditions that at $t=0$ and $t=T$, all the $x_i$'s are given. 

Note that we do not attempt to minimize the action, because it is minimal for a trajectory satisfying Newton's second law only for sufficiently short times~\cite{gray2007action}; in general, it is only a stationary point.

We replace the continuous time dependence with a discretization into $M$ time points. The time step
between adjacent time points is $\Delta t \equiv T/M$. Eq. \ref{eq:NewtonsLaw} is replaced
with a discretized second derivative

\[
   \label{eq:DiscreteNewtonsLaw}
   D^2 x_i = F_i(\{x_j\})
\]
where
\[
   \label{eq:DefOfD2}
    D^2 x(m) = \frac{x(m+1)-2x(m)+x(m-1)}{\Delta t^2}.
\]
We will minimize the objective function
\[
   \label{eq:DefOfObjective}
   O \equiv \sum_{m=1}^M \sum_{i=1}^{Nd} [ D^2 x_i(m) - F_i(\{x_j(m)\})]^2.
\]
There are two methods for doing this. The first is to directly apply the 
Fletcher-Reeves conjugate gradient algorithm ~\cite{fletcher1964function} directly
to these $(M-2)d N$ degrees of freedom.
Because of numerical rounding in double precision, this by itself is often not enough
to achieve a well converged minimum.
To overcome this numerical impediment, we devise a hybrid method similar to the shooting
method but iterative. 

\begin{figure}[htb]
\begin{center}
\includegraphics[width=0.4\hsize]{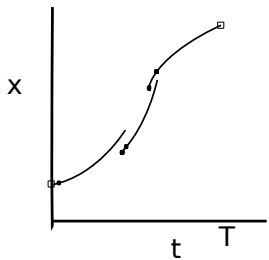}
\caption
{ 
An illustration of the hybrid shooting, and minimization procedure employed. The black dots
represent the time slice for the degrees of freedom being minimized. The number of time points
in each segment is $M_S$.
}
\label{fig:slice_shooting}
\end{center}
\end{figure}

We consider iterating  Eq.~\ref{eq:DiscreteNewtonsLaw} in time. Given $x_i(m)$ and $x_i(m-1)$,
this equation allows us to determine $x_i(m+1)$. This is known as the Newton-St\"ormer-Verlet
method~\cite{hairer2003geometric}. We also divide up the complete time interval of $M$ points into {\em segments}, each of length $M_S$. And we only iterate along individual segments. 
We iterate this discretization of Newton's law for $M_S$ time slices starting
from the first two adjacent time slices in a particular segment $S$. Therefore instead of having
the original number of variables to minimize, we have many less, because the only degrees of freedom
are the points that are initially given, as all other points are determined by iteration. For every $M_S$ time points, we attempt to match
the two starting points of a segment, with the two final points of the previous segment.
This is illustrated in Fig. \ref{fig:slice_shooting}. The black dots at the beginning
of a segment represent the degrees of freedom that will be adjusted so that the beginning
and end of overlapping adjacent segments agree. To make the notation less cumbersome,
we will suppress the particle indices, and write the coordinates of a segment starting at time slice $K$
as $[x(K),x(K+1),\dots,x(K+M_S),x(K+M_S+1)$.  All points are generated deterministically from the
initial two time points, $x(K)$ and $x(K+1)$ iterating Eq. \ref{eq:DiscreteNewtonsLaw}.
The adjacent segment starts $M_S$ time slices 
later and its coordinates are labeled  $[y(K+M_S),y(K+M_S+1),\dots,y(K+2M_S),y(K+2M_S+1)$.
To get these segments to match, we would like $x(K+M) = y(K+M)$ and $x(K+M+1) = y(K+M+1)$.
To achieve this iteratively with only two separate segments, we minimize some combination of quadratic
differences
\begin{multline}
   \label{eq:O_K}
   O_K(x,y) \equiv\\
   (x(K+M) - y(K+M))^2 + (x(K+M+1) - y(K+M+1))^2 \\
   +\lambda [(y(K+M+1)-y(K+M))-\\
   (x(K+M+1)-x(K+M))]^2  
\end{multline}
where $\lambda$ is a constant that determines how much weight is put in matching derivatives as
opposed to function values. It could be set to zero, but it was found that convergence was faster
with non-zero values, such as $\lambda = 10$. 

With multiple segments we minimize the objective function for this hybrid shooting method $O_{hybrid}$
\[
   O_{hybrid} \equiv \sum_{K} O_K
\]
using the same Fletcher Reeves conjugate gradient algorithm as above~\cite{fletcher1964function}.

As with the local minimization method, there is a limit to the precision of trajectories. To
improve on the minimum found, we use the final path obtained with segments of length $M_S$
as starting points for a new minimization iteration, where we double the path length of each segment from $M_S$ to $2 M_S$.
Now the number of points where matching is required is halved. This continues until the segment
is the full number of time slices in the system.

\section{Calculation cluster entropy}
\label{app:calc_fm}

\begin{figure}[htb]
\begin{center}
\includegraphics[width=0.4\hsize]{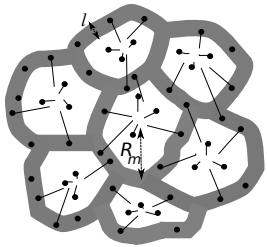}
\caption
{ 
A representation of a multi-clump trajectory where the clumps of size $R_M$ are larger than the mean free path
$l$. Particles starting deep inside of a clump must stay in it, but particles within a distance $l$
of the surface can choose to go to one of two clumps.
}
\label{fig:surfaces}
\end{center}
\end{figure}

We depict a multi-clump trajectory in Fig. \ref{fig:surfaces}. The thick gray lines represent the
surface regions between neighboring clumps. The width of these regions is the mean free path $l$. 
Particles that are initially in the grey regions are able to travel to either side of the surface,
meaning that they have a choice of which clump to join. Particles deep inside a clump, that is,
not in the grey regions, must stay inside their initial region traveling towards the center. 
The linear dimension $R_M$ of a clump is related to the number of particles in a clump $M$, through the number
density $n$ and the volume $\propto R_M^d$, implying that $R_M \sim (M/n)^{1/d}$.
Here we are considering  $M$ sufficiently large so that $R_M > l$.

The number of particles in the surface region involves its thickness $l$, its surface area, and the
density. Therefore for a single clump, the number of particles in this surface is $\sim n l R_M^{d-1} = n l (M/n)^{(d-1)/d}$.
The total number of particles in all surface regions (all of the grey area in Fig.  \ref{fig:surfaces})
is $N/M$ times this quantity, which is $(N/M) n l R_M^{d-1} = N l (M/n)^{-1/d}$.
Each particle in this region
has two choices of where to go. Therefore the entropy of these surface regions is $~\sim N l (M/n)^{-1/d} \log 2$.
The scaling of this with $M$ and the total number of particles $N$, is $N/M^{1/d}$. As expected, the
larger $M$, the smaller the surface area, and therefore the smaller this contribution. We should also include
the different configurations of these surfaces. For example if $d=2$, then the entropy of these
surfaces is bounded from above by assuming that surfaces are  random walks. In that case the surface
configurational entropy is also proportional to the surface area, giving an equal or a lesser
contribution to the entropy. The same statement should hold in higher dimension.
Therefore, taking into account the different surface configurations will not alter the dependence of the entropy of $M$. 

We conclude that as a function of $M$, the number of clusters $\mathcal{N_C}(M) \sim \exp(c N l (M/n)^{-1/d})$.
where $c$ is a constant. This decreases with increasing $M$ and is extensive in $N$, as claimed in
Sec. \ref{sec:ClusterAndTrajectories}.

\clearpage

\bibliography{classical_paths}

\end{document}